\documentclass[epjCONF]{svjour}
\usepackage{graphics}
\usepackage[varg]{txfonts} 
\usepackage[latin1]{inputenc}
\session-title{A Universe of dwarf galaxies: Observations, Theories, Simulations}
\begin{document}
\title{How non-linear scaling relations unify dwarf and giant elliptical galaxies}
\author{Alister W.\ Graham\fnmsep\thanks{\email{AGraham@swin.edu.au}} }
\institute{Centre for Astrophysics and Supercomputing, Swinburne University of Technology, Hawthorn, Victoria 3122, Australia.}

\abstract{ 
  Dwarf elliptical galaxies are frequently excluded from bright galaxy samples
  because they do not follow the same linear relations in diagrams
  involving effective half light radii $R_{\rm e}$ or mean effective
  surface brightnesses $\langle \mu \rangle_{\rm e}$.  However, using two linear
  relations which unite dwarf and bright elliptical galaxies we explain 
  how these lead to curved relations when one introduces either the 
  half light radius or the associated surface brightness.  In particular, the
  curved $\langle \mu \rangle_{\rm e}$-$R_{\rm e}$ relation is derived here.
  This and other previously misunderstood curved relations, once heralded as evidence for a
  discontinuity between faint and bright elliptical galaxies at $M_B \approx
  -18$ mag, actually support
  the unification of such galaxies as a single population whose structure
  (i.e.\ stellar concentration) varies continuously with stellar luminosity and mass.  }

\maketitle


\begin{figure*}
\begin{center}
\rotatebox{270}
{\resizebox{0.72\columnwidth}{!}{ 
\includegraphics{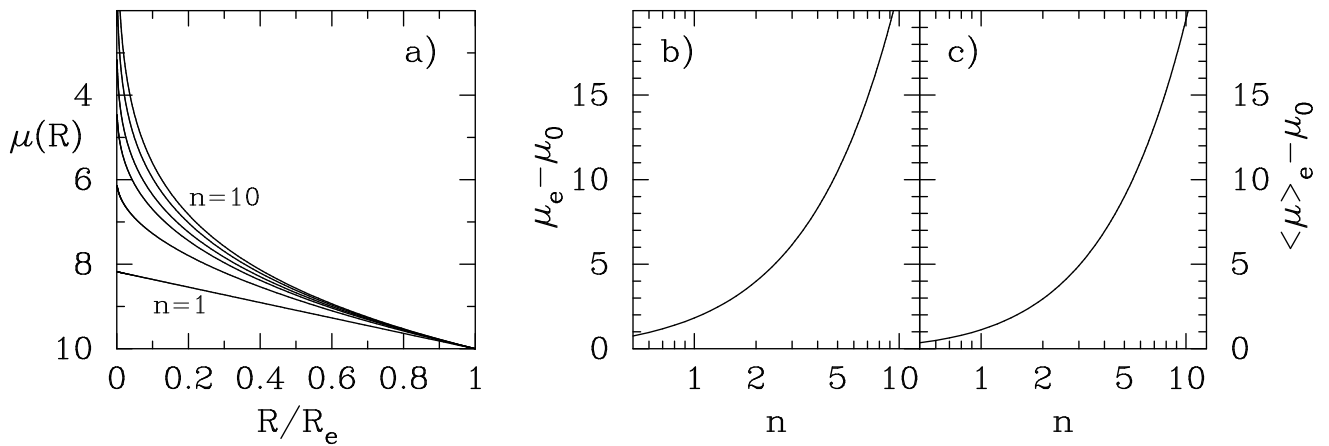} }}
\caption{Panel a) S\'ersic $R^{1/n}$ surface brightness profiles with
  effective surface brightness $\mu_{\rm e}$=10, and $n$=1, 2, 3, 4, 6 and 10.
Panels b) and c) show the difference between the central surface brightness at
  $R=0$, denoted by $\mu_0$, and the effective surface brightness $\mu_{\rm
  e}$ and the mean effective surface brightness $\langle \mu \rangle_{\rm e}$
  within the effective radius $R_{\rm e}$. 
}
\label{Fig1}
\end{center}
\end{figure*}

Elliptical galaxies, and the bulges of disc galaxies, do not have structural
homology (e.g.\ Davies et al.\ 1988; Caon et al.\ 1993; D'Onofrio et al.\
1994; Young \& Currie 1994,1995; Andredakis et al.\ 1995). 
Instead, they have a continuous range of stellar concentrations ---
quantified by the S\'ersic (1968) index $n$ (see Figure~\ref{Fig1}a) --- that
varies linearly with both stellar luminosity and central surface brightness
(after correcting for central stellar deficits or excess light).  An
unappreciated consequence of these two linear relations which unite faint and
bright elliptical galaxies across the alleged divide at $M_B \approx -18$ mag
is that relations involving either their effective half-light radius ($R_{\rm
e}$) or their effective surface brightness ($\mu_{\rm e}$), or the mean
surface brightness within $R_{\rm e}$ ($\langle \mu \rangle_{\rm e}$), will be
non-linear.
Such curved relations 
have often been heralded as evidence that different physical processes
must be operating on faint and bright elliptical galaxies 
because these relations have a different 
slope at either end.  
To further complicate matters, sample selection which
includes faint and bright elliptical galaxies, but excludes the
intermediate luminosity population, can effectively break such continuously
curved relations into two apparently disconnected relations. 

\begin{figure*}
\rotatebox{270}
{\resizebox{0.61\columnwidth}{!}{ 
\includegraphics{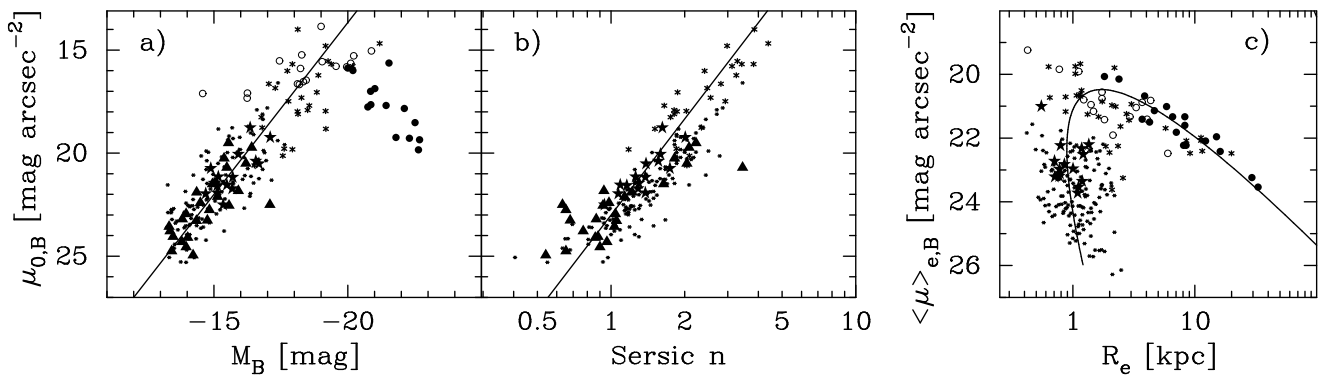} }}
\caption{Due to the observed linear relation of the B-band central surface
  brightness $\mu_{0,B}$ with a) the absolute magnitude $M_B$
  (Eq.~\ref{Eq_muM}) and b) the logarithm of the S\'ersic exponent $n$
  (Eq.~\ref{Eq_mun}), the relation between the effective radius $R_{\rm e}$
  and the mean surface brightness within this radius $\langle \mu
  \rangle_{\rm e}$ (Eq.~\ref{Eq_MuR}) is highly curved for elliptical 
  galaxies.  The somewhat orthogonal distribution in panel c) is not evidence
  for two different physical processes operating at the faint and bright end
  of the elliptical galaxy sequence.  Instead it is a consequence of the two
  linear relations which unify the faint and bright end, and bridge the
  alleged divide between dwarf and normal elliptical galaxies at $M_B \approx
  -18$ mag.  The ``core galaxies'' (large filled circles) with partially
  depleted cores can be seen to have lower central surface brightnesses than
  the relation in panel a).  However, the inward extrapolation of their outer
  profile yields $\mu_0$ values which follow the linear relation, as first
  noted by Jerjen \& Binggeli (1997).
  The data are from the compilation by Graham \&
  Guzm\'an (2003, their figure~9).  Dots represent dwarf elliptical (dE)
  galaxies from Binggeli \& Jerjen (1998), triangles represent dE galaxies
  from Stiavelli et al.\ (2001), large stars represent Graham \& Guzm\'an's
  (2003) Coma dE galaxies, asterix represent intermediate to bright E galaxies
  from Caon et al.\ (1993) and D'Onofrio et al.\ (1994), open circles
  represent the so-called ``power-law'' E galaxies from Faber et al.\ (1997),
  and the filled circles represent the ``core'' E galaxies from these same
  authors.  The S0 galaxies are excluded, pending bulge/disc decompositions. 
}
\label{Fig2}
\end{figure*}

Figure~\ref{Fig2} shows three diagrams for elliptical galaxies, two with
linear relations that naturally explain the third panel's curved relation. The data have
been taken from the compilation by Graham \& Guzm\'an (2003), while the two linear
relations from that paper have been slightly tweaked here.  From the first relation
between central surface brightness $\mu_0$ and S\'ersic index $n$, given by 
\begin{equation}
\mu_0 = 23 - 15.5\log(n)
\label{Eq_mun}
\end{equation}
and shown in Figure~\ref{Fig2}b, one can convert $\mu_0$ into
$\langle \mu \rangle_{\rm e}$ using the S\'ersic formula 
\begin{equation}
\langle \mu \rangle_{\rm e} = \mu_0 + 2.5b/\ln(10) -
2.5\log(n{\rm e}^b\Gamma(2n)/b^{2n}), 
\end{equation}
where $b \approx 1.9992n-0.3271$ 
(Figure~\ref{Fig1}c, see Capaccioli 1989 and Graham \& Driver 2005, their equations 7 and 9). 
The associated effective radius $R_{\rm e}$ is acquired by matching the second relation 
between absolute magnitude $M$ and central surface brightness, given by 
\begin{equation}
M = 0.6\mu_0-28.2  
\label{Eq_muM}
\end{equation}
and shown in Figure~\ref{Fig2}a, with the magnitude formula
\begin{equation}
M = \langle\mu\rangle_{\rm e} - 2.5\log(2\pi R_{\rm e,kpc}^2) - 36.57 
\end{equation}
(e.g.\ Graham \& Driver 2005, their equation~12).  Doing this yields the expression
\begin{equation}
\log R_{\rm e} = \frac{1}{5}\left\{ 0.4\langle \mu \rangle_{\rm e} +1.5\left[ \frac{b}{\ln(10)}
  -\log\left(\frac{ne^b\Gamma(2n)}{b^{2n}}\right) -6.91\right] \right\}, 
\label{Eq_MuR}
\end{equation}
in which one knows the value of $n$ associated with each value of $\langle \mu
\rangle_{\rm e}$ from the expressions above. 
Equation~\ref{Eq_MuR}, obtained from two empirical linear relations 
(equations~\ref{Eq_mun} and \ref{Eq_muM}), 
is a curved relation that is shown in Figure~\ref{Fig2}c.
%

The implications of this should not be glossed over.  Without any
understanding of the $\langle\mu\rangle_{\rm e}$-$R_{\rm e}$ diagram
(Figure~\ref{Fig2}c), it has in the past been used to claim that faint and
bright elliptical galaxies must have obtained their structure from different
physical processes --- because the faint and bright arms of the galaxy distribution
are nearly perpendicular to each other.  If there was instead one linear
relation in this diagram, it would have been claimed that a single unifying
mechanism was operating.
As seen in Figures~\ref{Fig2}a and \ref{Fig2}b, linear relations do however exist across the
faint and bright end of the galaxy distribution in $M$-$\mu_0$ and $n$-$\mu_0$
space.  In passing we note that because of 
the linear relation between $M$ $(=\log L)$ and $\log n$ (e.g.\ 
Caon et al.\ 1993; Young \& Currie 1995; Jerjen \& Binggeli 1997; 
Graham, Trujillo \& Caon 2001; Ferrarese et al.\ 2006), and 
the associated non-linear behaviour between $\mu_0$ and
$\mu_{\rm e}$ (Figure~\ref{Fig1}b), 
the relation between $M$ and $\mu_{\rm e}$ is not linear.  Similarly, as
detailed above, the relation between $\langle\mu\rangle_{\rm e}$ and $R_{\rm
  e}$ is not linear but curved.  This result, however, 
has just been predicted/explained 
from linear relations which unify faint and bright elliptical galaxies.

For those curious about the discrepant ``core galaxies'' in
Figure~\ref{Fig2}a, this paragraph and the following one provide something of
an explanatory detour.
The departure from the $B$-band $M_B$-$\mu_{0,B}$ diagram by elliptical galaxies
brighter than $M_B \approx -20.5$ mag (Mass $> 0.5$-$1\times 10^{11} M_{\odot}$), 
seen in Figure~\ref{Fig2}a, was explained 
by Graham \& Guzm\'an (2003) in terms of partially depleted cores
{\it relative to their outer S\'ersic profile} (see also 
Graham 2004; Trujillo et al.\ 2004; Merritt \& Milosavljevi\'c 2005).  
Such cores are thought to have formed 
from dry galaxy merger events (Begelman, Blandford, \& Rees 1980; Ebisuzaki,
Makino, \& Okumura 1991) and resulted in Graham et al.\ (2003) 
and Trujillo et al.\ (2004) advocating 
a ``new elliptical galaxy paradigm'' based on the presence of this central
stellar deficit versus either none or an excess of light (see also 
Gavazzi et al.\ 2005; Ferrarese et al.\ 2006; C\^ot\'e et al.\ 2007; 
and later Kormendy et al.\ 2009). 
As discussed in Graham \& Guzm\'an (2003), this distinction at $M_B \approx 
-20.5$ mag between elliptical galaxies with partially depleted cores and those without, 
is a separate issue from the alleged division between elliptical galaxies and
dwarf elliptical galaxies at $M_B \approx -18$ mag (e.g.\ Kormendy 1985).

Further evidence for this division at $M_B \approx -20.5$ mag,
rather than any division at $M_B \approx -18$ mag 
(Kormendy 1985; Kormendy et al.\ 2009; Tolstoy et al.\ 2009), 
mag comes from the tendency for 
the brighter elliptical galaxies to be anisotropically pressure supported
systems with boxy isophotes, while the less luminous early-type galaxies 
are reported to have discy isophotes and often contain a rotating disc (e.g.\ Carter 1978, 
1987; Bender 1988; Peletier et al.\ 1990; Emsellem et al.\ 2007; Krajnovi\'c
et al.\ 2008).
Additional support for the above mentioned dry merging scenario at the high-mass end is
the flattening of the colour-magnitude relation above 0.5-1$\times 10^{11}
M_{\odot}$.
As discussed by Graham (2008, his section~6) and reiterated by
Bernardi et al.\ (2010), this flattening was evident in 
Baldry et al.\ (2004, their Figure~9) and Ferrarese et al.\ (2006, their
Figure~123), and even Metcalfe, Godwin \& Peach (1994). 
This flattening has since been shown in other data sets (e.g., Skelton, Bell \&
Somerville 2009, although they reported the transition at $M_R=-21$ mag, 
i.e.\ $\approx$1 mag fainter).  Finally, 
the change in slope of the luminosity-(velocity dispersion) relation is also 
supportive of a transition at around $M_B \approx -20.5$ mag (e.g.\ Davies et al.\
1983; Held et al.\ 1992; De Rijcke et al. 2005; Matkovi\'c \& Guzm\'an 2005). 
%
%
In passing, it is remarked that the $L$-$\sigma$ 
relation would be steeper at the high-luminosity end if total, luminosity-weighted, 
infinite aperture velocity dispersions (equal to one-third of the virial
velocity dispersion) were used instead of central velocity dispersions, 
as required in equation~1 from Wolf et al.\ (2010)
and advocated by many papers in 1997 for use in the ``Fundamental Plane'' (FP: 
Djorgovski \& Davis 1987). 

Returning to Figure~\ref{Fig2}, 
due to the linear relations in Figure~\ref{Fig2}a and \ref{Fig2}b 
which connect dwarf and ordinary elliptical
galaxies across the alleged divide at $M_B \approx -18$ mag, 
coupled with the smoothly varying change in light profile shape as a function
of galaxy magnitude, the $\langle \mu \rangle_{\rm e}$-$R_{\rm e}$ relation is
curved.  The apparent deviant nature of the dwarf elliptical galaxies from the
approximately linear section of the bright-end of the $\langle \mu
\rangle_{\rm e}$-$R_{\rm e}$ distribution, known as the Kormendy (1977)
relation, does not imply that two different physical processes are operating.

Similarly, the location of disc galaxy bulges at the faint end of this
distribution does not imply that they must be ``pseudobulges''.  That is, 
``pseudobulges'', as opposed to ``classical bulges'', 
can not be identified simply because they are outliers from 
the Kormendy (1977) relation (Gadotti 2009) --- the bright arm of a longer, 
continuous and unifying curved relation. 
While such apparent outliers are associated with bulges having low luminosities, 
low S\'ersic indices, and faint central surface brightnesses, 
this is not by itself evidence that they experienced a different formation process. 
For similar reasons, galaxies which do not follow the bright arm of the curved
$L$-$R_{\rm e}$ relation (derived/explained in Graham \& Worley 2008, their figure~11) need
not be pseudobulges, nor are galaxies which do not follow the bright arm of
the curved Mass-$R_{\rm e}$ relation (presented by Graham et al.\ 2006, their figure~1b). 
Galaxies which do not follow the bright arm of the continuous, but curved,
$L$-$\langle \mu \rangle_{\rm e}$ and $L$-$\mu_{\rm e}$ 
relation (e.g.,\ Graham \& Guzm\'an 2003,
their Figure~12) also need not necessarily be pseudobulges (Greene, Ho \&
Barth 2008; Fisher \& Drory 2010). 

While luminous bulges and elliptical galaxies follow the same Fundamental
Plane (Falc\'on-Barroso, Peletier \& Balcells
2002), fainter elliptical galaxies and bulges smoothly depart from the FP
(when sample selection biases do not chop out a gulf between the faint and bright
systems).  These systems appear to follow a continuous trend along what is a
curved manifold, of which the FP is the flat portion of this curved
hypersurface (Graham \& Guzm\'an 2004; Graham 2005; La Barbera et al.\ 2005;
Zaritsky, Gonzalez \& Zabludoff 2006; Gargiulo et al.\ 2009).  The curved
nature of this manifold can be derived from a number of linear relations which
span and unify faint and bright elliptical galaxies, and it implies that some
over-riding physical process dictates their structure.  Galaxies which appear
to branch off from the faint end of the Fundamental need not have formed from
different physical mechanisms.

In summary, using curved relations, that can be constructed from unifying linear
relations, as a means to identify an allegedly different class of galaxy
(i.e.\ dwarf elliptical galaxies or pseudobulges) is not appropriate.  
The curved relations involving either $R_{\rm e}$ or $\langle \mu \rangle_{\rm
  e}$, and also $\mu_{\rm e}$ (see Figure~\ref{Fig1}),
do not signal a different formation mechanism for low- and high-luminosity
elliptical galaxies.  Instead, these curved relations can be understood in terms of, and indeed
predicted from, linear relations known to unify faint and bright elliptical
galaxies.  Understanding the implications of structural non-homology (i.e.\
the range of stellar concentrations) among elliptical galaxies (and bulges in
disc galaxies) is key to better understanding galaxies and the
connections they share.

\acknowledgement

For those who attended this talk, they will know that while the above material 
was presented, it was not the original nor primary subject matter. 
For readers interested in the coexistence of nuclear star clusters (NC) and massive
black holes (BH), and the nature of the smooth transition from NC-dominance to
BH-dominance as the host bulge or elliptical galaxy mass increases, I refer
one to Graham \& Spitler (2009) and Graham et al.\ (2010).


\begin{thebibliography}{}
 \bibitem{APB95}Andredakis Y.C., Peletier R.F., Balcells M., MNRAS \textbf{275}, (1995) 874
 \bibitem{Bald4}Baldry I.K., Glazebrook K., Brinkmann J., Ivezi{\'c} {\v Z}., Lupton R.H., Nichol R.C., Szalay A.S., ApJ \textbf{600}, (2004) 681
 \bibitem{BBR80}Begelman M.C., Blandford R.D., Rees M.J., Nature \textbf{287}, (1980) 307
 \bibitem{Ben88}Bender R., A\&A \textbf{193}, (1988) L7
 \bibitem{Berna}Bernardi M., Roche N., Shankar F., Sheth R.K., MNRAS, submitted (2010, arXiv:1005.3770)
 \bibitem{BaJ98}Binggeli B., Jerjen H., A\&A \textbf{333}, (1998) 17
 \bibitem{CCD93}Caon N., Capaccioli M., D'Onofrio M., MNRAS \textbf{265}, (1993) 1013
 \bibitem{Cap89}Capaccioli M., \textit{The World of Galaxies}, ed. H. G. Corwin, L. Bottinelli (Berlin: Springer-Verlag 1989) 208
 \bibitem{Car78}Carter D., MNRAS \textbf{182}, (1978) 797
 \bibitem{Car87}Carter D., ApJ \textbf{312}, (1987) 514
 \bibitem{Cote7}C{\^o}t{\'e} P., et al., ApJ \textbf{671}, (2007) 1456
 \bibitem{Dav88}Davies J.I., Phillipps S., Cawson M.G.M., Disney M.J., Kibblewhite E.J., MNRAS \textbf{232}, (1988) 239
 \bibitem{Dav83}Davies R.L., Efstathiou G., Fall S.M., Illingworth G., Schechter P.L., ApJ \textbf{266}, (1983) 41
 \bibitem{Sven5}de Rijcke S., Michielsen D., Dejonghe H., Zeilinger W.W., Hau G.K.T., A\&A \textbf{438}, (2005) 491
 \bibitem{DaD87}Djorgovski S., Davis M., ApJ \textbf{313}, (1987) 59
 \bibitem{DCC94}D'Onofrio M., Capaccioli M., Caon N., MNRAS \textbf{271}, (1994) 523
 \bibitem{EMO91}Ebisuzaki T., Makino J., Okumura S.K., Nature \textbf{354}, (1991) 212
 \bibitem{Ems07}Emsellem E., et al., MNRAS \textbf{379}, (2007) 401
 \bibitem{Fab97}Faber S.M., et al., AJ \textbf{114}, (1997) 1771 
 \bibitem{Falc2}Falc\'on-Barroso J., Peletier R.F., Balcells M., MNRAS \textbf{335}, (2002) 741 
 \bibitem{Fet6a}Ferrarese L., et al., ApJS \textbf{164}, (2006) 334
 \bibitem{FaD10}Fisher D.B., Drory N., ApJ \textbf{716}, (2010) 942
 \bibitem{Gadot}Gadotti D.A., MNRAS \textbf{393}, (2009) 1531
 \bibitem{Garg9}Gargiulo A., et al., MNRAS \textbf{397}, (2009) 75
 \bibitem{Gav05}Gavazzi G., Donati A., Cucciati O., Sabatini S., Boselli A., Davies J., Zibetti S., A\&A \textbf{430}, (2005) 411 
 \bibitem{Gra04}Graham A.W., ApJ \textbf{613}, (2004) L33
 \bibitem{Gra05}Graham A.W., IAU Colloquia 198, ``Near-Field Cosmology with Dwarf Elliptical Galaxies'', H. Jerjen \& B. Binggeli (eds.), (Cambridge, Cambridge University Press, 2005) 303-310
 \bibitem{Gra8a}Graham A.W., ApJ \textbf{680}, (2008) 143
 \bibitem{GaD05}Graham A.W., Driver S.P., PASA \textbf{22(2)}, (2005) 118
 \bibitem{GETAR}Graham A.W., Erwin P., Trujillo I., Asensio Ramos A., AJ \textbf{127} (2003) 1917
 \bibitem{GaG03}Graham A.W., Guzm\'an R., AJ \textbf{125}, (2003) 2936
 \bibitem{GMMJT}Graham A.W., Merritt D., Moore B., Diemand J., Terzi{\'c} B., AJ \textbf{132}, (2006) 2711 
 \bibitem{GOnke}Graham A.W., Onken C.A., Athanassoula E., Combes F., MNRAS,
  submitted (2010, arXiv:1007.3834)
 \bibitem{GaS09}Graham A.W., Spitler L.R., MNRAS \textbf{397} (2009) 2148
 \bibitem{GTC01}Graham A.W., Trujillo I., Caon N., AJ \textbf{122} (2001) 1707
 \bibitem{GaW08}Graham A.W., Worley C.C., MNRAS \textbf{388}, (2008) 1708
 \bibitem{GHB08}Greene J.E., Ho L.C., Barth A.J., ApJ \textbf{688}, (2008) 159
 \bibitem{Hel92}Held E.V., de Zeeuw T., Mould J., Picard A., AJ \textbf{103}, (1992) 851
 \bibitem{JaB97}Jerjen H., Binggeli B., \textit{The Nature of Elliptical Galaxies; The Second Stromlo Symposium} (ASP Conf.\ Ser. 1997), v.116, p.239
 \bibitem{Kor77}Kormendy J., ApJ \textbf{218}, (1977) 333
 \bibitem{Kor85}Kormendy J., ApJ \textbf{295}, (1985) 73
 \bibitem{KFCB9}Kormendy J., Fisher D.B., Cornell M.E., Bender R., ApJS \textbf{182}, (2009) 216
 \bibitem{Kraj8}Krajnovi{\'c} D., et al., MNRAS, \textbf{390} (2008) 93
 \bibitem{Barb5}La Barbera F., Covone G., Busarello G., Capaccioli M., Haines C.P., Mercurio A., Merluzzi P., MNRAS \textbf{358}, (2005) 1116
 \bibitem{MaG05}Matkovi\'c A., Guzm\'an R., MNRAS \textbf{362}, (2005) 289
 \bibitem{MaM05}Merritt D., Milosavljevi{\'c} M., Living Reviews in Relativity \textbf{8}, (2005) 8
 \bibitem{MGP94}Metcalfe N., Godwin J.G., Peach J.V., MNRAS \textbf{267}, (1994) 431
 \bibitem{PDIDC}Peletier R.F., Davies R.L., Illingworth G.D., Davis L.E., Cawson M., AJ \textbf{100}, (1990) 1091
 \bibitem{Ser68}S\'ersic J.-L., Atlas de Galaxias Australes (Cordoba: Observatorio Astronomico, 1968)
 \bibitem{SBS09}Skelton R.E., Bell E.F., Somerville R.S., ApJ \textbf{699}, (2009) L9
 \bibitem{Sti01}Stiavelli M., Miller B.W., Ferguson H.C., Mack J., Whitmore B.C., Lotz J.M., AJ \textbf{121}, (2001) 1385 
 \bibitem{THT09}Tolstoy E., Hill V., Tosi M., 2009, ARA\&A, 47, 371
 \bibitem{Truj4}Trujillo I., Erwin P., Asensio Ramos A., Graham A.W., AJ \textbf{127}, (2004) 1917
 \bibitem{Wolf0}Wolf J., Martinez G.D., Bullock J.S., Kaplinghat M., Geha M., Mu{\~n}oz R.R., Simon J.D., Avedo F.F., MNRAS \textbf{406}, (2010) 1220
 \bibitem{YaC94}Young C.K., Currie M.J., MNRAS \textbf{268}, (1994) L11
 \bibitem{YaC95}Young C.K., Currie M.J., MNRAS \textbf{273}, (1995) 1141
 \bibitem{ZGZ06}Zaritsky D., Gonzalez A.H., Zabludoff A.I., ApJ \textbf{638}, (2006) 725
\end{thebibliography}
\end{document}